# BAKTRAK: Backtracking drifting objects using an iterative algorithm with a forward trajectory model[1]

Øyvind Breivik[+,*], Tor Christian Bekkvik[**], Cecilie Wettre[*] and Atle Ommundsen[#]

[+]Corresponding author. E-mail oyvind.breivik@met.no
[*]Norwegian Meteorological Institute, Alleg 70, NO-5007 Bergen, Norway / Geophysical Institute, University of Bergen
[**]Christian Michelsen Research, Bergen, Norway
[#]Norwegian Defence Research Establishment, Kjeller, Norway

## Abstract

The task of determining the origin of a drifting object after it has been located is highly complex due to the uncertainties in drift properties and environmental forcing (wind, waves and surface currents). Usually the origin is inferred by running a trajectory model (stochastic or deterministic) in reverse. However, this approach has some severe drawbacks, most notably the fact that many drifting objects go through nonlinear state changes underway (e.g., evaporating oil or a capsizing lifeboat). This makes it difficult to naively construct a reverse-time trajectory model which realistically predicts the earliest possible time the object may have started drifting. We propose instead a different approach where the original (forward) trajectory model is kept unaltered while an iterative seeding and selection process allows us to retain only those particles that end up within a certain time-space radius of the observation. An iterative refinement process named BAKTRAK is employed where those trajectories that do not make it to the goal are rejected and new trajectories are spawned from successful trajectories. This allows the model to be run in the forward direction to determine the point of origin of a drifting object. The method is demonstrated using the Leeway stochastic trajectory model for drifting objects due to its relative simplicity and the practical importance of being able to identify the origin of drifting objects. However, the methodology is general and even more applicable to oil drift trajectories, drifting ships and hazardous material that exhibit non-linear state changes such as evaporation, chemical weathering, capsizing or swamping. The backtracking method is tested against the drift trajectory of a life raft and is shown to predict closely the initial release position of the raft and its subsequent trajectory.

## 1 Introduction

Modeling the fate and trajectory of drifting objects or the drift and weathering of hydrocarbons on the ocean surface is a demanding task that puts every part of an operational model suite to the test as it requires high-quality wind fields, current fields and possibly also wave fields (Hackett *et al*, 2006, Davidson *et al*, 2009). Objects and hydrocarbons have very specific drift and weathering properties of which little may be known, and in real situations even the last known position (LKP) may be uncertain. The errors in wind, wave and current fields will further add to the uncertainties of the trajectory forecasts.

---

The task of backtracking from a known position to the time and position to where the object started drifting is an even more challenging task. In principle a trajectory model may be run backwards if the horizontal forcing fields are sufficiently non-divergent and if the properties of the object do not undergo non-linear state changes such as swamping, capsizing, sinking, or drowning in the case of drifting objects, or weathering, evaporation and vertical mixing in the case of hydrocarbons. This straight forward implementation of backtracking is for example employed by Ambjörn (2008) in an operational oil drift backtracking service for the Baltic Sea. The condition of nearly divergence-free flow is usually met in large-scale geophysical flows, but nonlinear and step-like state changes in the properties of the advected matter (be it oil, algae or solid objects) are impossible to handle simply by reversing the direction of the wind field and the current field.

There exists a large number of *forward* oceanic trajectory models which model passive tracers based on single-model current fields or weighted combinations of heterogeneous models (Rixen and Ferreira-Coelho, 2007; Rixen *et al*, 2008; Vandenbulcke *et al*, 2009). Trajectory models also exist for a range of more complex drifting entities, such as oil (Hackett *et al*, 2006; Hackett *et al*, 2008) and search and rescue objects (Breivik and Allen, 2008; Davidson *et al*, 2009). However, unlike for the atmosphere, where backtracking pollution over large distances in the planetary boundary layer using both forward and backward (adjoint) trajectory models is an established scientific discipline (Rao *et al*, 2007; Stohl 1996; Stohl 2002), tracking drifting objects and substances to their origin in the ocean is still a fledgling field. Reverse-time models of passive tracers exist (see e.g. Callies *et al*, 2011) and have been used for tracking nearshore sources of hazardous materials to their origins (Havens *et al*, 2009). However, backtracking more complex objects or substances with nonlinear properties in reverse time requires an adjoint, and this may not be possible to establish. Especially when dealing with hydrocarbons with nonlinear weathering processes it becomes crucial to assess whether a particular spill or object may indeed have come from a suspected release location and release time.

Here we explore an new approach to traditional reverse-time backtracking where a stochastic oceanic trajectory model is initialized and run in the forward direction, whereupon the individual ensemble members (hereafter referred to as particles) that come within an acceptable time-space distance of the observation are used to initialize a new forward integration. Unsuccessful particle trajectories are discarded. This procedure is then iterated until an acceptable number of trajectories end up within the target area (defined as a time-space radius around the observations location) is reached and a time-space distribution of possible initial locations for the drifting object has emerged.

The method is conceptually similar to sequential importance resampling (Doucet *et al*, 2001), a statistical filtering method which has been proposed as a data assimilation technique in geophysical systems (see Van Leeuwen, 2009, and references therein), but has so far not been tested on oceanic trajectory integrations. However, there are also important differences between sequential importance resampling and our iterative backtracking procedure. Where particle filters in geophysical systems typically deal with modest ensembles of large-dimensional numerical models and a large number of observations, trajectory models have a small-dimensional state-space (the two-dimensional position vector of the particles), a relatively large ensemble, and usually just one observation (the final location of the object, known here as the target area).

The forward iterative method has not been tried before on oceanic trajectory models. The objective of this work is to outline the procedure and to test it on a weakly nonlinear stochastic trajectory model, the Leeway model (Breivik and Allen, 2008), an ensemble trajectory model for estimating search areas for drifting objects. The choice of trajectory model is motivated by the immediate societal value of backtracking objects to their origin, but also by the relative simplicity of the trajectory model. Though the Leeway model is only weakly nonlinear, it is a useful test bed for the method because we have field data of a drifting object under controlled conditions to compare with. It is important to note that the methodology is equally applicable to more complex trajectory models with for example hydrocarbon chemistry and weathering, but exactly because of the simplicity of the Leeway model we argue that it is easier to gauge the efficacy of the iterative procedure here than on more sophisticated trajectory models.

This article is organized as follows. Section 2 describes the iterative method and the forward trajectory model and its forcing fields (surface current and wind). Then the results from a set of integrations are described in Section 3 before a discussion of the convergence and the potential pitfalls of the method are discussed in Section 4. Finally, Section 5 concludes and points the way forward to other implementations of the method.

# 2 The BAKTRAK method

The description of the method divides naturally into two parts; (i) the forward trajectory model (Section 2.1) and (ii) the iterative refinement of the initial conditions that eventually lead to a set of trajectories that end up at the right time in the target area (Section 2.2). We refer to the procedure as BAKTRAK, but the actual forward model can be arbitrarily chosen, as can the forcing fields employed.

## 2.1 Leeway, an ensemble trajectory model for drifting objects

The Leeway model is a stochastic ensemble trajectory model for drifting objects. The model has been described in detail by Breivik and Allen (2008). Here we give only a brief recount of its main features. We follow Allen and Plourde (1999) and Breivik *et al* (2011) and define the leeway as the motion of the object induced by wind (10 m reference height) and waves relative to the ambient current (between 0.3 and 1.0 m depth). The model computes the trajectories of a large number of particles with stochastic perturbations that move under the influence of wind and current. The leeway of a particular object is empirically estimated, usually from open ocean field trials (see Breivik *et al*, 2011).

### 2.1.1 Modeling the leeway (windage) of a drifting object

The leeway vector is commonly decomposed into a downwind and crosswind component. Leeway field trials are required to determine the relation between the wind speed and the leeway speed and divergence angle, $L$ and $L_a$, or more robustly, the downwind (DWL) and crosswind (CWL) leeway components, $L_d$ and $L_c$. The objects are assumed to respond linearly to changes in the wind speed (Breivik and Allen, 2008). Objects on the sea surface will drift at an angle relative to the downwind vector given by the ratio of CWL to DWL (the leeway divergence angle), either to the left or to the right of the downwind vector. Whether an object has oriented itself to the left or to the right is usually unknown and both eventualities must be accounted for by the ensemble. This often leads to a search area consisting of two high-probability areas. However, objects may also *jibe*, i.e., abruptly shift from a left to a right tack. The model assumes a 4% per hour probability that a given particle shifts orientation relative to the downwind vector. This makes the model slightly nonlinear and will over time (in the forward sense) "fill in" areas between the two high-probability parts of the search area.

The empirical relation between leeway and wind speed is discussed in more detail by Allen and Plourde (1999), Allen (2005), Breivik and Allen (2008) and Breivik *et al* (2011). In its current operational manifestation, some 63 classes of drifting objects described by Allen and Plourde (1999) plus object classes more recently explored (see Allen *et al*, 2010; Breivik *et al*, 2011 and Turner *et al*, 2009) are available.

### 2.1.2 Ensemble of perturbed trajectories

Small perturbations are added to the wind field and the leeway parameters of individual particles to account for the uncertainties in the empirical estimates of the leeway of the drifting object under consideration (Breivik *et al*, 2011). The perturbations of the wind vectors are time-dependent (new perturbations are added to each time step) to simulate high-frequency fluctuations in the wind field, while the perturbations to the leeway parameters are constant in time to simulate the different loading conditions of an object class. The trajectories are computed using a regular second order Runge-Kutta method with wind field perturbations added to the first (trial) step of the Runge-Kutta integration.

### 2.1.3 Coastlines and stranding
The current implementation of the Leeway model utilizes the Global, Selfconsistent, Hierarchical, High-resolution Shoreline (GSHHS) coastline contour (Wessel and Smith, 1996). The use of a high-resolution coastline contour allows accurate determination of stranding of particles. However, the ocean model grid has a resolution of 4 km in our operational implemenation, and to make the model capable of handling near-shore conditions, current vectors are extrapolated toward the coastline. Particles "stick" to the shoreline and can not detach themselves once stranded.

### 2.1.4 Model domain
The model has been set up to cover the North Sea, the Norwegian Sea and the Barents Sea with 10-m wind vectors from a 12 km resolution numerical weather prediction model, the High-resolution limited area model (HIRLAM, see Undén *et al*, 2002) and surface current fields from a 4 km resolution ocean model (Engedahl, 1995; 2001), but the model domain is only restricted by the availability of wind and current fields (see Figure 1). The ocean model is run twice daily out to 60 hours. A seven-day rolling archive of two-hourly temporal resolution is continually updated. The model ingests GRIB version 1 fields in polar stereographic, rotated spherical and plate carrée (longitude-latitude) projections. In the standard *forward* setup (integration from a last known position), the model releases $O(500)$ particles in an area and over a period of time believed to enclose the incident (where the object started drifting). However, in the iterative backtracking model a much larger amount of particles is required as only a small fraction can be expected to reach the target area.

## 2.2 Iterative backtracking through seeding and selective breeding

The iterative backtracking procedure encapsulates the stochastic trajectory model described in the previous section. The model is run forward, but by discarding trajectories that do not end up near the target area an iterative refinement of the initial conditions is achieved. The successful trajectories are used to seed the next forward integration, but small perturbations are added to the initial time and location of the new particles. In general, it may not be known how far back to go, but often it is known. We impose a backwards time-limit of 7 days based on the availability of our rolling archive of current and wind. In cases where the *time* of the incident is known (if for example a distress call has been picked up) we can also start the integration at a given time and release all particles at that particular instant.

### 2.2.1 Initial seeding
The stochastic trajectory model is initialized with a fixed number, $O(5000)$, of initial particle positions drawn from a two dimensional circular normal (Gaussian in both longitude and latitude) distribution based on a first guess radius. The radius is estimated from a coarse upper bound on the speed of the drifting object, assumed here to be 2 m/s. The initial time of release of the particles is drawn from a uniform distribution. The ensemble of particles is thus spread over a range of initial positions and start times ranging from the earliest time that the object is assumed to have started drifting up until the time of observation in the target area. The trajectory model is now run forward from the earliest time that the object may have started drifting while particles are continually seeded in a smaller and smaller disc centered on the target area.

### 2.2.2 Selection of parent (breeder) particles by relative distance metric

In general, particles released near the target area in time and space will have a higher probability of reaching the target than "older" particles starting further away. Because both young and old particles should be given an equal chance of being selected for feedback, older particles are retained if they fall within a greater radius of the observation location than "younger" particles. Particles are thus considered for selection based on a metric denoted the *relative distance*,

$$\beta = \frac{D_1}{D_0} \quad (1),$$

where $D_0$ is the particle's initial distance to the target area and $D_1$ its distance at the end of the integration. Another argument for using relative distance as a selection criterion for the next generation of particles is that the computational cost of requiring a given number of particles to hit an absolute target area increases quickly with increasing distance from target. By allowing a larger target area for particles starting further away the computational cost can be maintained at tolerable levels.

At the end of the first iteration, no particles may yet have arrived inside the target area radius. But the backtrack feedback algorithm still needs some initial particles for seeding the next iteration. Also, when in later iterations more particles successfully reach the target area, not all are needed in the feedback loop as the seeder generates replicas of successful particles for the next iteration. Therefore, the $N_p$ (default set to 64) nearest particles are used in the feedback loop even though they initially may arrive far outside TA.

Particles which arrive further away from TA may also be used for the seeding of the next iteration in cases where few particles get close to TA, but their starting positions used for the resampling are adjusted for original distance offset from TA to allow for faster convergence. This procedure is also self-correcting in that it will also increase the initial seed distance from target area in cases where the assumed upper bound on the speed of the drifting object has been set too low.

Note that the spatial resolution of the forcing fields (in our case 4 km resolution surface current and 12 km resolution 10-m wind vector fields) in itself does not affect the convergence of the method. In fact, a higher resolution current vector field will normally disperse nearby particles more efficiently due to small-scale eddies. The procedure is more lenient on particles that have drifted over large distances as it considers the particle's relative distance, hence even in a complex current field some of these particles will be used to seed the next iteration.

### 2.2.3 Seeding the next iteration

The $N_p$ "parent" particles that have been deemed successful are used as the starting point of a new forward integration. The $N_c$ "children" particles seeded from each selected parent particle of the previous run are spread out according to a radial Gaussian distribution about their parent particle according to the initial distance of their parent particle to TA,

$$r = sD_0 \quad (2),$$

where $s$ is a dimensionless spread factor (default 0.1) and $r$ is the seed radius (equal to twice the radial standard deviation, $\sigma = r/2$) about the parent particle. The spread factor ensures that parent particles that started far away (large $D_0$) from TA are replicated by a wide cloud of "offspring" particles while parent particles that originated near TA are replicated by a tight cloud. A similar spread in initial time is also implemented. The seeder thus regenerates an ensemble of size $N_p N_c \sim O(5000)$ particles from perturbations to the initial time and location of these parent particles and the next integration is started.

### 2.2.4 Halting condition

The iteration stops when the number of successful particles no longer continues to rise and/or when a sufficient number of particles, $O(100)$, have come within a critical time-space radius of TA. What is considered a "sufficient" number of particles is somewhat subjective, but the procedure is flexible and easily altered. The Leeway model is not idempotent, *i.e.*, rerunning the model ensemble never produces the exact same solution due to stochastic perturbations both to the initial locations and the wind field. This means that no amount of further refinement of the initial conditions can increase the number of particles that will reach the target area. Hence the only way to increase the number of successful particles is to boost the number of initial particles, *N*. The iterative algorithm is summarized in Table 1.

# 3 Life raft drift experiment

A field campaign was conducted with the Norwegian Coast Guard vessel CGV Ålesund outside the island of Fedje on the west coast of Norway from 21-30 March 2011 (around 60º40'N, 004º20'E, see Figure 2). The objective of the cruise was to establish the drift properties of search and rescue objects, among them a Viking 12 person life raft. The raft was loaded to typical conditions with sandbags (approximately 400 kg) and equipped with a tracking transponder, a Global Position System (GPS) data logger and a 2 m high WeatherPak wind anemometer mast. The raft was deployed at 2011-03-22T18:20 UTC and picked up after approximately 14 hours at 2011-03-23T08:30 UTC. The wind was south-westerly and westerly 5-15 m/s and the current was mostly northbound 0.3-0.8 m/s (see Figure 3). The area is monitored with high frequency (HF) radars (see Breivik *et al*, 2011, Breivik and Sætra, 2001 and Essen *et al*, 2003). There was reasonable agreement between the ocean model and the HF current fields during the campaign (see Figures 2 and 3).

We have compared the BAKTRAK trajectory ensemble with the true trajectory of the Viking life raft. The leeway properties of the raft were assumed similar to that of the shallow ballast rafts studied by Allen and Plourde (1999). We assumed mean values of a canopied life raft (see Table 2).

Figure 4 illustrates the initial seeding described in Section 2.2.1. The model is seeded with *O*(5000) particles (the default) over a period of 16 hours from 2011-03-22T16 UTC until 07 UTC the following day (an hour before pickup of the raft). The seeding is Gaussian in both longitude and latitude, as shown in panels (b) and (c). A large part of the particle trajectories end on the shoreline (black dots in Figure 4, panel (a) are particles that were seeded over land). Unsuccessful particle trajectories are shown in gray in Figure 4, panel (d). The filtering procedure itself does not distinguish between stranded particles and particles still drifting but simply selects successful particles based on their proximity in time and space to the target area.

The first iteration assumes that the object has traveled at a maximum radial speed toward TA of 2 m/s. The particles are thus spread according to their initial release time (the further back in time the wider the radius) and spreads particles according to a two-dimensional Gaussian distribution. Panels (b) and (c) of Figure 4 show the particle distributions in latitude and longitude at a given time. Panel (d) shows the successful trajectories in red against the backdrop of *all* trajectories (gray). Those trajectories that came nearest TA in terms of the relative distance of Eq (1) were selected for breeding the next generation.

The next stage is to select at most $N_p$ from those successful trajectories and replicate each as $N_c$ "children". The "parent" initial locations and release times with small perturbations provide the starting point for the particles of the next iteration. This procedure, as described above, goes on until a satisfactory number of trajectories reach their destination. Panel (a) of Figure 5 shows the new initial distribution based on the successful particles of the first iteration, two hours after the first particles were released. Panel (b) shows the trajectories of the total ensemble in gray and the trajectories selected for the final iteration in red. It is evident from panels (a) and (b) that after the initial integration quite a range of initial locations vie for selection. As the iterations continue, the initial locations near the final position are weeded out in favor for the earlier initial locations, until in panel (c) only those locations surrounding the time and position of the true release position have won out. Note also that the spread factor *s* in Eq (2) is used to spread the "children" around their parent particle. The further away from TA the parent particle is (large $D_0$), the larger the spread. This effect is apparent in panel (a) of Figure 5, where patchy clouds of offspring particles of various sizes according to their distance $D_0$ to TA are present. As the iterations progress, this field of initial particle locations is refined until the increase in successful particles slows and starts to vacillate about a convergence limit. Panel (c) of Figure 5 shows the initial field of particles at the start of iteration 7 together with the observed trajectory of the life raft. Now the ensemble of initial locations has converged toward a much tighter cloud surrounding the drop location of the raft. Panel (d) shows the trajectories of the total ensemble in gray and the trajectories selected for the final iteration in red.

Figure 6 shows iteration no 8, chosen as the final even though nine iterations were performed (the integration with the most successful particles is selected as the final one). The convex hull of the initial locations at the time of release of the life raft is shown in blue in the south-western part of Figure 6. The release position is enclosed by the convex hull (note that a few particles were released before or after the time of the release of the life raft and their initial location falls outside the convex hull). As time progresses, the real trajectory enters the center of the ensemble (shown as red trajectories), until the final pickup location is found in the center of the convex hull (blue) of ensemble.

# 4 Convergence of the BAKTRAK procedure

A series of BAKTRAK integrations have been carried out to test the convergence under a realistic range of coastal and offshore weather and current conditions. Their convergence is listed in Figure 7. We measure convergence in terms of "successful" particles of the final iteration. The success or failure of a particle is decided by its relative distance as defined by Eq (1), i.e., a particle from afar can end up further from the target area than a particle that started closer and still be selected for breeding. This makes sense when iterating as it allows particles further afield to compete with those nearer in space and time to the target area. It also offers a convenient objective measure of convergence when we compare different integrations with different environmental conditions and different integration times. However, it is not strictly necessary to adhere to the concept of relative distance when we present the final integration, as we are really interested in the *absolute* distance to the target area when we present the final backtracked ensemble. However, the question of what is considered close enough crops up when using an absolute measure of distance to the target area. As an example, consider Figure 8 where a threshold distance of 3000 m to target area has been subjectively chosen. This is slightly larger than the radius of the convex hull in Figure 6 (2.3 km), but the ensemble of successful particles is more than doubled to 542. Because of the level of subjectivity involved in choosing an acceptable target area radius, we stick to the automated convergence estimated from the relative distance estimates, noting that these convergence rates are lower bounds on the convergence in real world applications where the user could subjectively select the target area radius.

The model was tested on three nearshore cases, named Sheringham Shoal, Shetland and F/V Idarson (a real case with a fishing vessel where nothing was known about its distress location). The locations of all test cases are marked in Figure 1. All three cases showed acceptable convergence rates in the presence of blocking by land, similar to what was seen for the life raft study (although with onshore winds the effect is less pronounced). The filtering procedure does not distinguish between stranded particles and particles still drifting but simply selects successful particles based on their proximity in time and space to the target area. The convergence is acceptable in all three cases. The time range from earliest to latest possible time of release varied from five hours to three days.

The Ekofisk integration is set in the middle of the North Sea to explore how realistic wind and current conditions affect the convergence of the iteration over longer periods in offshore conditions. Particles are released over 24 hours in a relatively large radius about the target area at Ekofisk, located at 56°30'N, 003°12'E. The currents are moderate and tidally dominated while the wind is south-westerly 7-10 m/s. Figure 7 shows the convergence to be somewhat slower than for the other integrations. The wide spread of the unsuccessful trajectories (not shown) even after eight iterations clearly shows that the effect of the open ocean is to slow the convergence as a much wider area of potential origins must be considered than in cases like the life raft study where a relatively straight coastline blocks one half-plane of the initial locations.

The life raft case study is quite representative of a nearshore accident and a sighting of the vessel a few hours later. The model shows reasonable convergence in the presence of complex coastlines and with variable current and wind conditions. Convergence on a larger time scale has been tested for the Ekofisk case, and even longer time scales (three days) for the Shetland case. In all cases, convergence of at least $O(100)$ particles can be achieved with 5-8 iterations on runs with $O(5000)$ particles. This number can be increased arbitrarily by expanding the size of the ensemble.

Although the examples here have presented the iteration with the largest number of particles that converge on the target area in this final forward integration, a simple way to boost the number of successful trajectories is to aggregate trajectories from previous iterations. This increases the tally typically by a factor five. The convergence is also seen to be quite rapid, although significant deviations occur from one iteration to the next. The convergence is not monotonous, but this is partly due to the flexibility in defining "successful" trajectories selected for breeding. In the early iterations (1 and 2), the successful trajectories may actually end quite far from the target area and still be selected for breeding. This allows the iterative procedure to adjust the initial distance in cases where the particles have been seeded too near the target area. The procedure usually reaches a maximum after 5-7 iterations, after which nothing more is added. The runs selected as "final" are the ones with the highest number of particles arriving at the target area.

As with reverse-time trajectory models, the question of whether the final set of backtracked trajectories is close to the true trajectory or not depends on the quality of the modeled current and wind vector fields as well as our ability to properly account for the object's behavior (the leeway and jibing in the case of our trajectory model). Other things being equal, it is clear that the reverse-time approach can generate a much larger ensemble and for cases where a simple passive tracer is sufficient, the reverse-time approach is clearly recommended. However, the BAKTRAK methodology described here is *not* intended for passive tracers but rather for complex trajectory models where a simple reverse-time is difficult to implement. In the case of the raft described in Section 3 the sensitivity to the ocean model is relatively small as the wind forces on its over-water structure dominate its drift. In such a case the pertinent question is whether the backtracking model (be it a reverse-time or an iterative forward method) properly accounts for all the forces on the drifting object. In other cases where the wind is weaker, a simple reverse-time model may well do a good job, but in general it is difficult to decide in advance whether a reverse-time trajectory computation will be sufficient.

# 5 Conclusion and perspectives for iterative forward trajectory models

We have established that the iterative method works for a weakly nonlinear stochastic trajectory model with complex geography and high-resolution forcing fields (wind and current). The model has been implemented on the same model domain as the operational Leeway model, covering the North Sea, the Norwegian Sea and the Barents Sea. The model backtracks objects to seven days ago. The BAKTRAK procedure circumvents the need for constructing a reverse-time version (adjoint) of the trajectory model, which may be difficult given the nonlinearities of model. This has the advantage of utilizing the full nonlinear properties of the trajectory model to determine whether a certain source location is a realistic one, but comes at the cost of a somewhat higher numerical cost and time of execution. The iterative procedure is self-correcting in the sense that larger initial distributions in time and space are tried for the next iteration if too few successful trajectories make it to the target area. The convergence toward an acceptable number of trajectories is shown to be quite rapid, and 5-8 iterations are sufficient for the implementation presented here. This has been confirmed through a range of integrations under different topographic constraints and in varying weather and current fields. Successful trajectories from subsequent iterations can also be aggregated to boost the ensemble, and as discussed in Section 4, the target area radius can also be subjectively increased or decreased for real world cases.

BAKTRAK reverse drift estimates showed close agreement with the observed trajectory of a life raft released during a field campaign in the Norwegian Coastal Current. Although more field work is clearly required to quantify the uncertainties under a range of weather conditions, the example clearly confirms the approach is valid and usable. Although beyond the scope of this work, it might be possible to construct an adjoint to the Leeway model where the weakly nonlinear effect of the jibing is accounted for. This would provide an interesting benchmark for the forward iterative method, but it might be difficult to disentangle the effect of the jibing on the adjoint. However, for more complex models such as oil drift such an adjoint may be very difficult to construct.

Future work should explore the potential of using the iterative approach outlined here for more complex trajectory models involving strong nonlinearities, in particular oil spill models (see Hackett *et al*, 2006; Broström *et al*, 2008; Hackett *et al,* 2009). A common approach to identifying spills is to investigate their chemical composition and compare with the chemical fingerprint of known sites (Christensen *et al*, 2004). Our method should prove useful as a complementary line of investigation when identifying sources of spills, whether from ships, land-based production plants (see Havens *et al*, 2009) or offshore oil fields.

The dynamical risk assessment analysis explored by Eide *et al* (2007) estimates the time to shore of ships following a planned route under realistic weather conditions. A backtracking algorithm applied to the ship drift model employed by Eide *et al* (2007) could explore the corresponding risk of objects running into fixed installations near a sea route.

Another line of investigation is to utilize graphics processing units (GPU) for massively parallel computational problems. This would allow a radical expansion of ensemble size to further refine the initial conditions (Brodtkorb *et al*, 2010). This method could also be used to investigate the much larger ensembles required to recreate observed *trajectories* rather than single observations (in our work denoted the target area*)*. One application where observed trajectories could be used would be in iceberg modeling. Recreating the iceberg trajectories using a forward procedure similar to what has been described here would require a very large number of trajectories and starts to resemble particle filtering in that more than one observation is available (van Leeuwen, 2009).

Another approach would be to employ forcing fields from different model systems, possibly consisting of both ocean models, wave models and numerical weather prediction models and weighing the different fields using the hyper ensemble techniques described by Rixen and Coelho (2007), Rixen *et al* (2008) and Vandenbulcke *et al* (2009). The iterative refinement technique described here is equally applicable to fields thus generated and would complement these techniques by allowing in principle any forward trajectory model to be utilized. In conclusion, the advantages of retaining the full forward trajectory model should more than outweigh the higher computational cost associated with the iterative BAKTRAK procedure when considering more complex trajectory models.

# Acknowledgments

This work was made possible by funding from the Norwegian Defence Research Establishment (FFI) through the BAKTRAK project. The field work was organized and partly funded by the project "Uncontrolled drift of ships and larger objects", funded by the Research Council of Norway (NFR) under the MAROFF programme (grant no 200862). The analysis has also benefited from the MAROFF project FARGE (grant no 200843). The authors wish to thank the Joint Rescue Co-ordination Centres (JRCC) and the Norwegian Navy for their continued support of the development of operational trajectory models for search and rescue. This work builds on results from the SAR-DRIFT project under the French-Norwegian Foundation (Eureka grant E!3652) and the FOB project funded through the NFR MAROFF programme (grant no 180175). This project has also benefited from the Norwegian National Supercomputing Facility (NOTUR). Finally, we wish to thank the two anonymous reviewers for their thorough scrutiny of the manuscript which made us include important new material.

# List of figures

Figure 1. Excerpt of the polar stereographic model domain of the ocean model. The model has an approximate resolution of 4 km and is run twice daily to 60 hours. A seven-day rolling archive of two-hourly temporal resolution is continually updated. The operational implementation of the trajectory model ingests current and wind vectors in polar stereographic, plate carrée and rotated spherical projections. Only every sixth surface current vector is shown for clarity. The locations of the five test cases are marked with red crosses. From south west (left) to north east (right): Sheringham Shoal (53º00'N, 000º23'E), Ekofisk (56º30'N, 003º12'E), Shetland (60º53'N, 000º37'E), Fedje life raft (60º45'N, 004º20'E), and F/V Idarson (70º15'N, 021º15'E).

Figure 2. The trajectory of the life raft (deployed at the SW end of the trajectory) overlaid on the surface current vector field from a network of high-frequency radars valid at 2011-03-23T02 UTC. The Norwegian Coastal Current flows swiftly through the area with currents exceeding 1 m s$^{-1}$.

Figure 3. The surface current vector field (blue isolines in ms$^{-1}$) of the operational ocean model, Nordic4 (4 km resolution) and 10-m wind vectors (red wind barbs measured in knots) from the HIRLAM12 (12 km resolution) numerical weather prediction model valid at 2011-03-23T02 UTC, approximately halfway through the drift experiment. Black crosses mark the deployment and pickup locations for the life raft.

Figure 4. The first BAKTRAK iteration of the life raft experiment. Panel (a) shows the initial distribution of all particles two hours after release (red, with particles seeded on land in black). A blue cross marks the life raft pickup location, used as the center for the integration. Panels (b) and (c) show the initial Gaussian

distribution in latitude and longitude of the particles, respectively. Panel (d) shows the 20 successful trajectories in red against the backdrop of unsuccessful trajectories (gray) that did not make it to the target area.

Figure 5. The second and seventh BAKTRAK iterations. Panels (a) and (c) show the initial distribution of the second and seventh iterations (raft pickup location marked with a cross). Panels (b) and (d) show the successful trajectories of the second and seventh iterations in red against the backdrop of unsuccessful trajectories (gray). It is evident from a comparison of panels (a) and (c) that many of the initial locations selected for iteration 2 are since weeded out in favor of a much tighter initial distribution centered near the true release position of the life raft in panel (c). The observed trajectory of the life raft is overlaid on the initial distribution of iteration 7 in panel (c) for reference. Panel (d) shows a marked increase in successful trajectories from panel (b).

Figure 6. The eighth iteration was selected as the final integration because it had the highest number of successful trajectories (red). The convex hulls of the initial distribution and the final distribution of the particles are shown in blue. The life raft trajectory is shown as black circles. The initial distribution of the ensemble encloses the release position of the life raft.

Figure 7. The number of successful trajectories, i.e., those that appear within reasonable proximity of the target area, determined by the relative distance metric in Eq (1) is shown as a function of iteration number. The aggregate number of successful trajectories is shown in the legend. Note that this number can become much higher if the target area radius is chosen subjectively, as shown in Figure 8.

Figure 8. The target area radius can be chosen arbitrarily. This may increase the number of successful trajectories, as here where the target area radius has been set to 3000 m. This more than doubles the number of successful trajectories (see Figure 6). The convex hull of the final distribution (blue) is now nearly circular because of the way the successful trajectories have been chosen.

# List of tables

Table 1. The BAKTRAK algorithm for the refinement of the initial time and position of particles for the next forward integration of the trajectory model.

```
-------------------------------------
BAKTRAK iterative feedback algorithm
-------------------------------------
Variables:
TA: Target area (where particles shall end up)
T₀: First release time for particles
T₁: Final release time for particles
D₀: Distance to TA from the N particle positions at the end of the previous iteration
D₁: Distance to TA from the N particle positions at the start of the next iteration
β = D₁/D₀, relative distance
Nₚ: Parent particles, the nearest (default 64) particles(sorted by β) from the
      previous iteration
N_c: Children particles centered on a parent particle from previous iteration
N = Nₚ*N_c: Ensemble size
S: dimensionless spread factor (default 0.1)
R =S*D₀

Algorithm:
      1. First iteration:
            1.1 Seed N initial particles in radius R around TA over
                  period T₀-T₁
      2. Iterate until enough particles reach TA:
            2.1 Run forward trajectory model
            2.2 Sort particles by increasing relative distance, β
            2.3 Select Nₚ nearest (in terms of β) parent particles
            2.4 Seed N_c children particles from Gaussian distr with radius
                  R (2σ) around each of the Nₚ parent particles
```

Table 1. A summary of the BAKTRAK procedure.

| Object | Downwind leeway (DWL), $L_d$ | | | Crosswind leeway (CWL), $L_c$ | | |
|---|---|---|---|---|---|---|
| | Slope (%) | Offset (cm s⁻¹) | $S_{yx}$ (cm s⁻¹) | Slope (%) | Offset (cm s⁻¹) | $S_{yx}$ (cm s⁻¹) |
| Life raft shallow ballast, mean values | 2.7 | 0.0 | 12.0 | 1.1 | 0.0 | 9.4 |

Table 2. The leeway coefficients used in the life raft drift experiment. We have assumed the mean values for shallow ballast life rafts (see Allen (2005). The standard deviation $S_{yx}$ is used to spread the leeway coefficients of the ensemble.

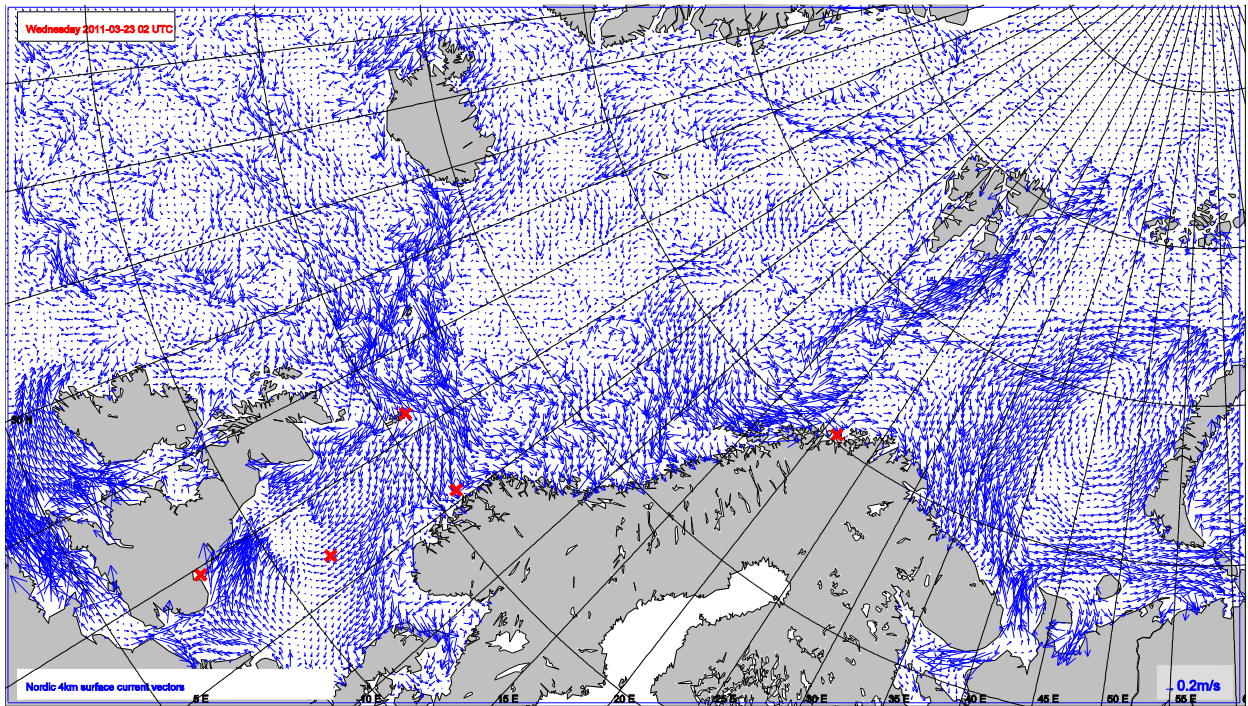

Figure 1.

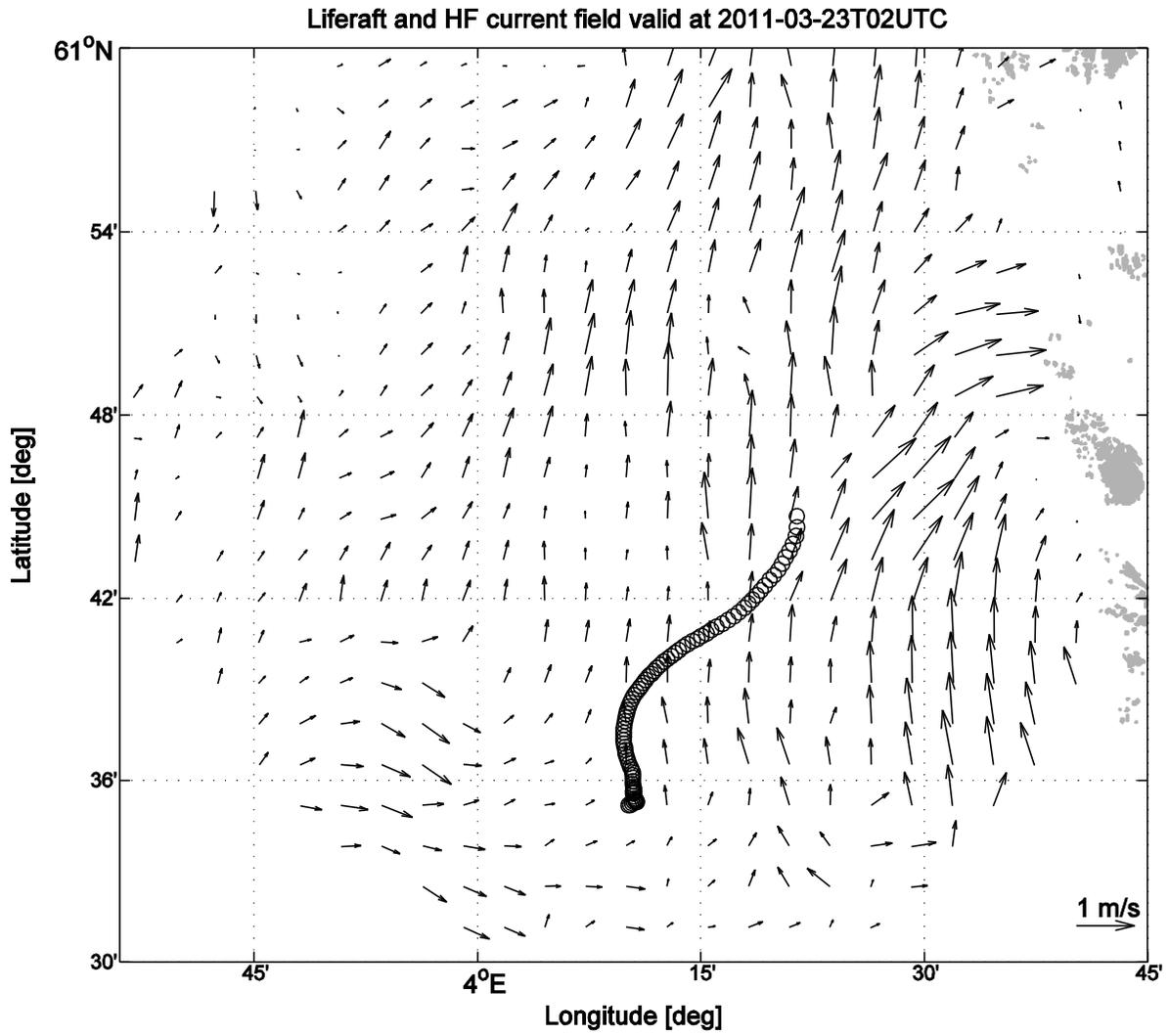

Figure 2

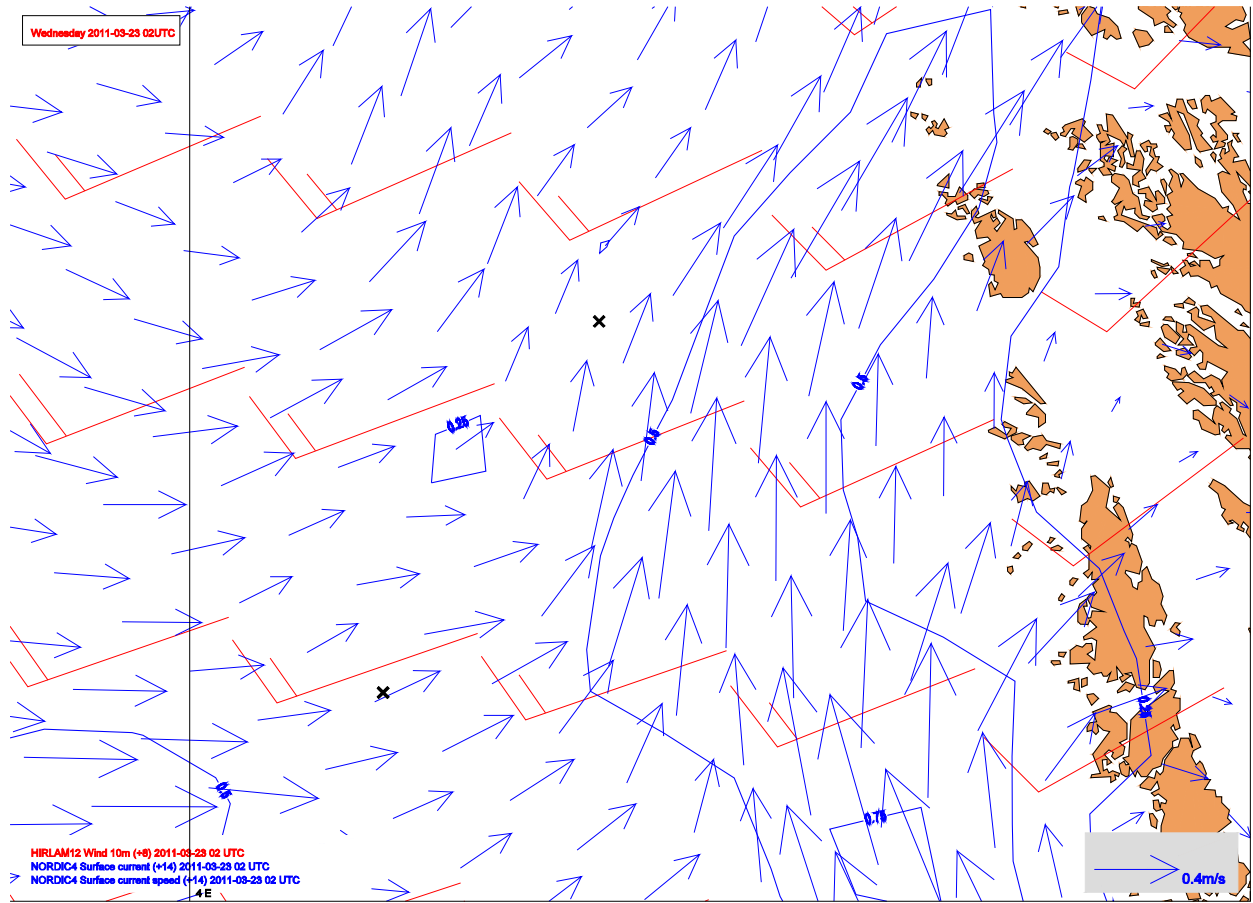

Figure 3.

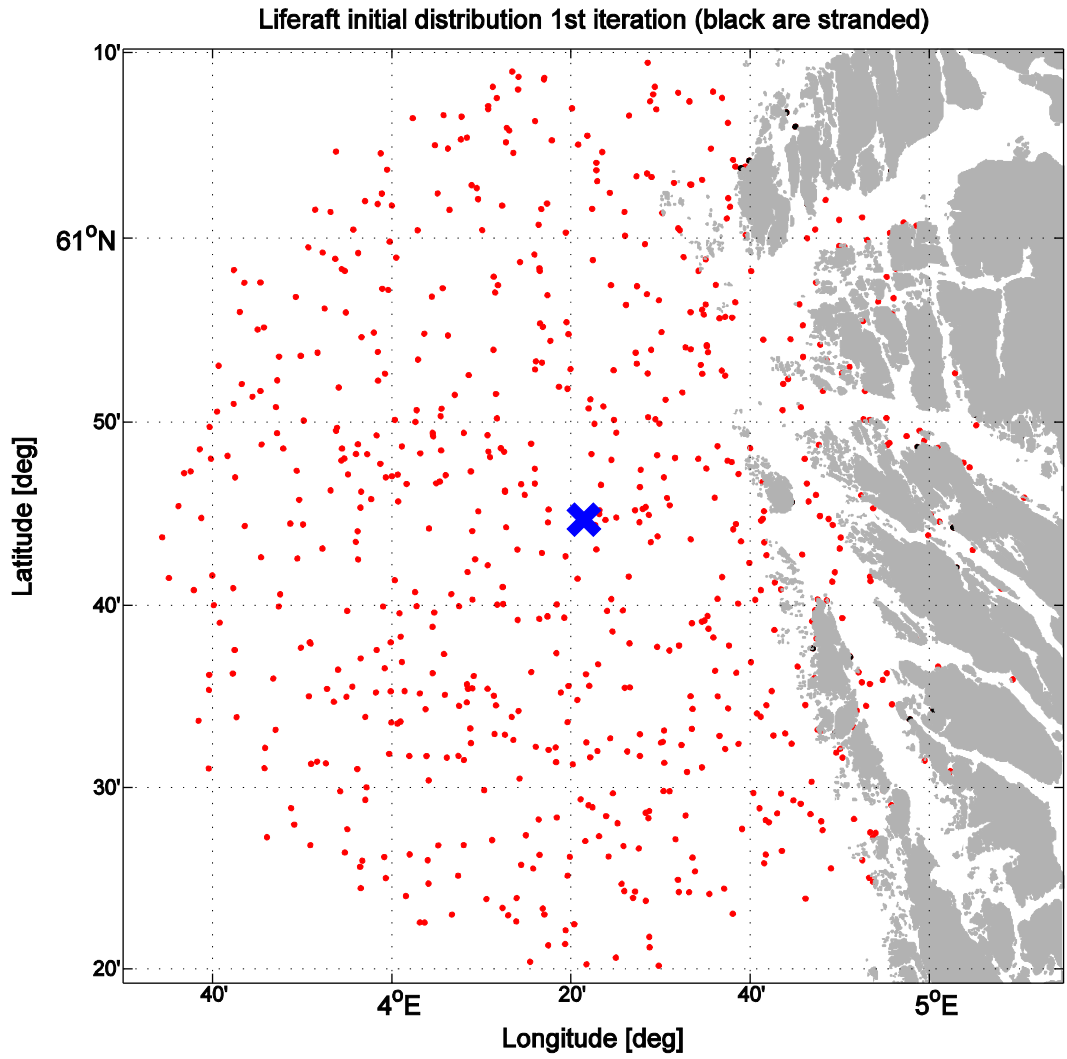

Figure 4, panel (a).

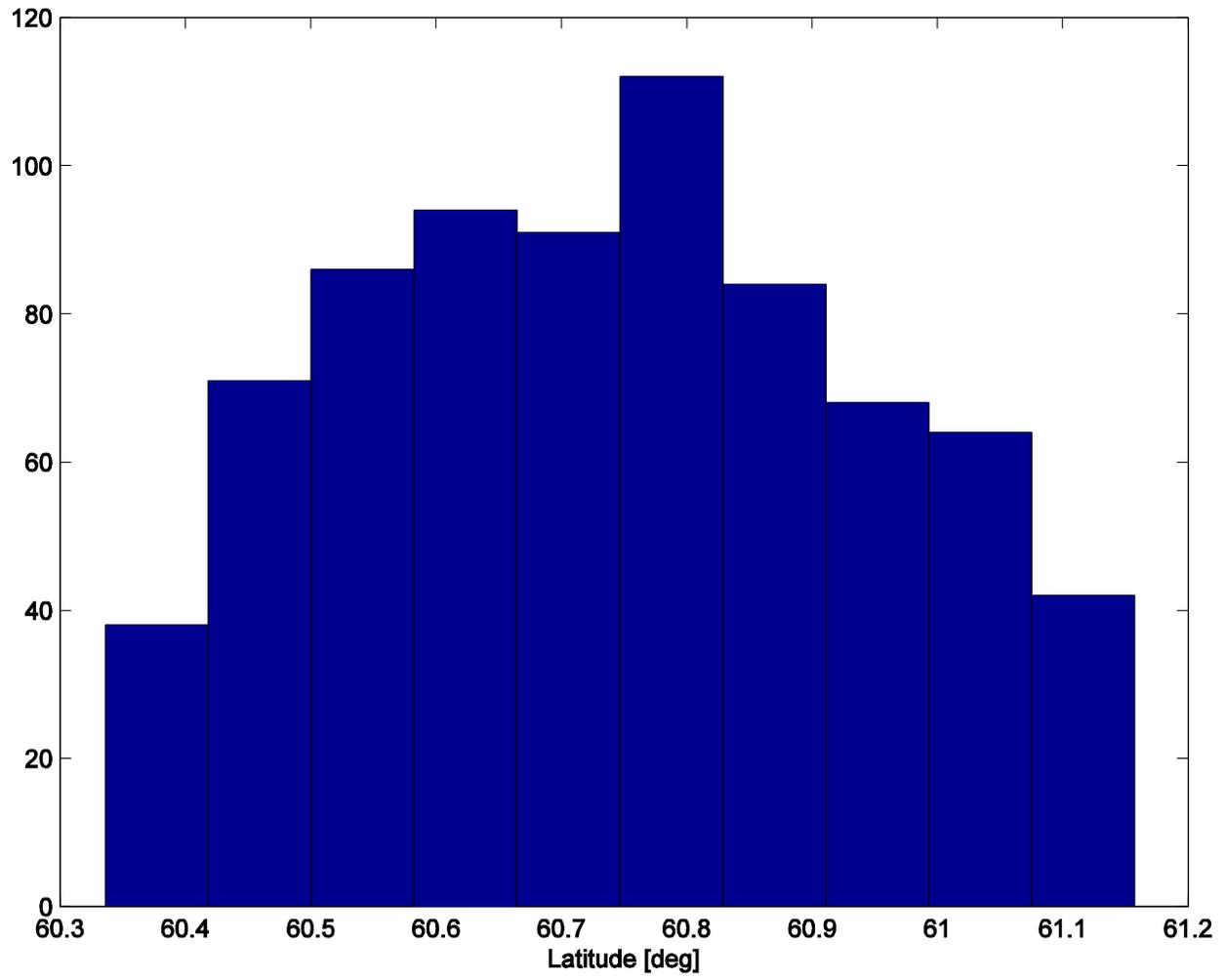

Figure 4, panel (b).

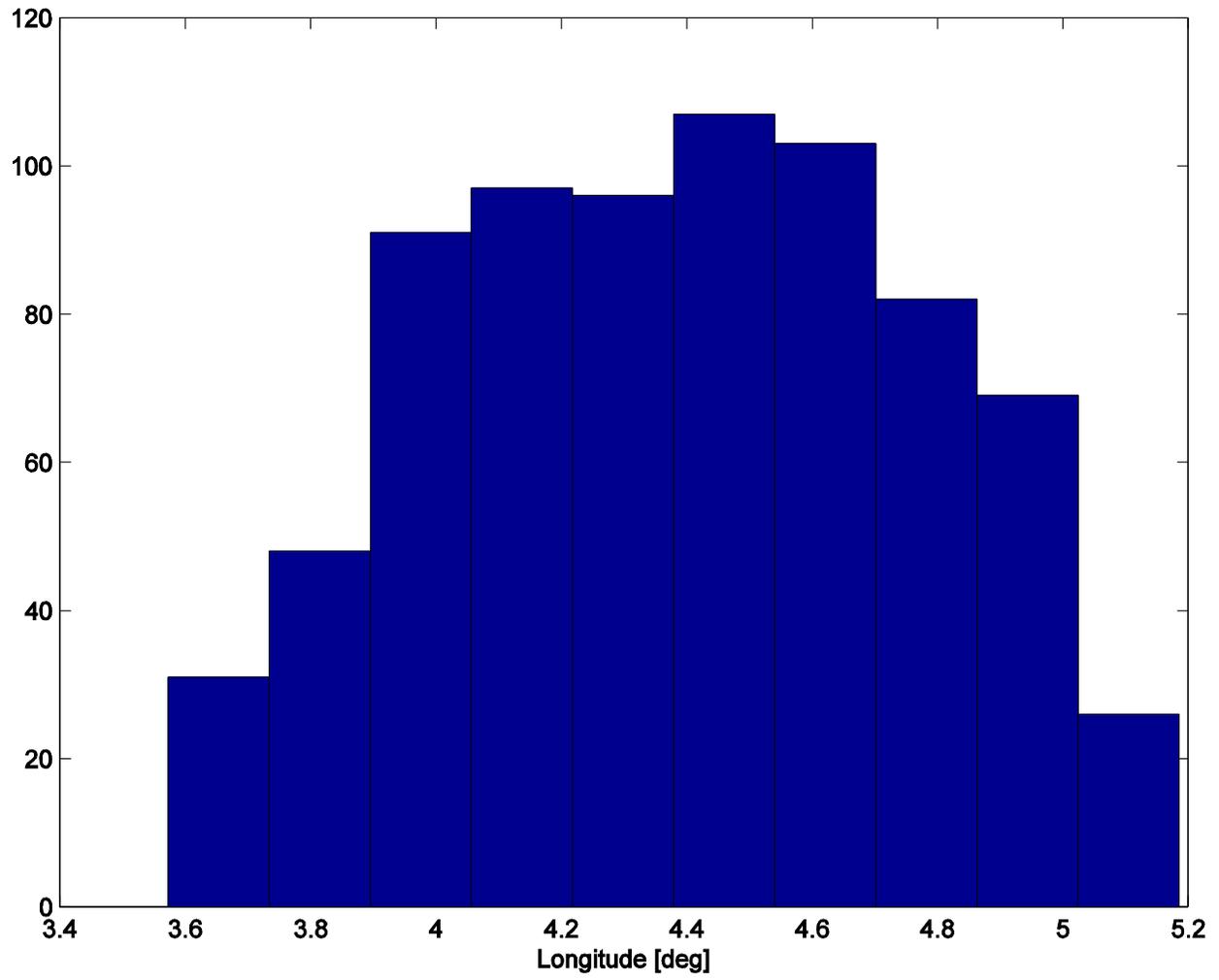

Figure 4, panel (c).

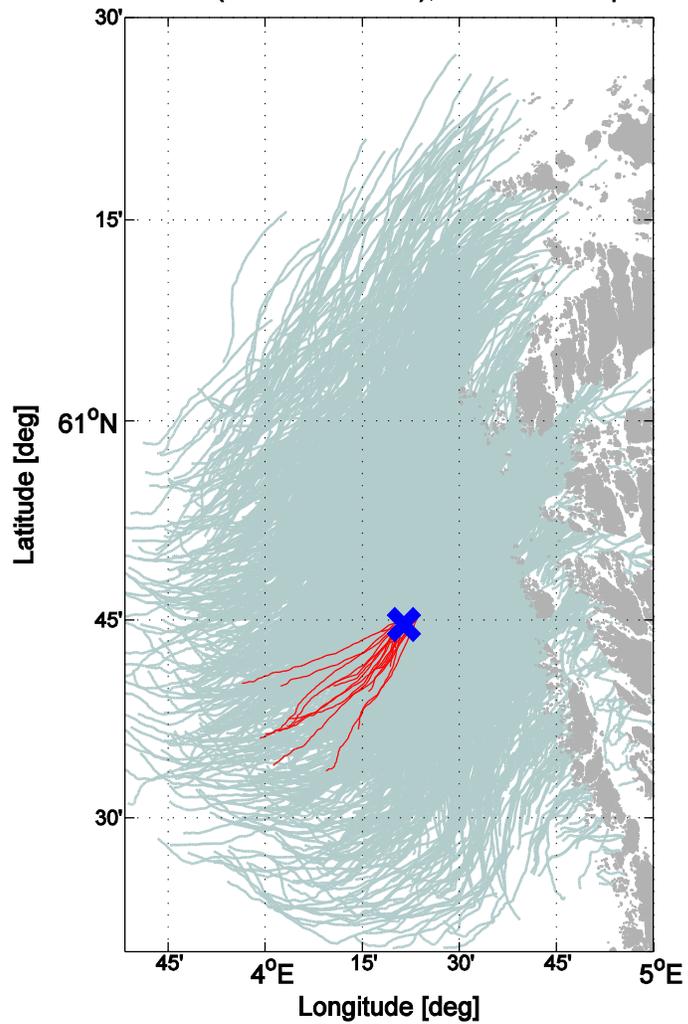

Figure 4, panel (d).

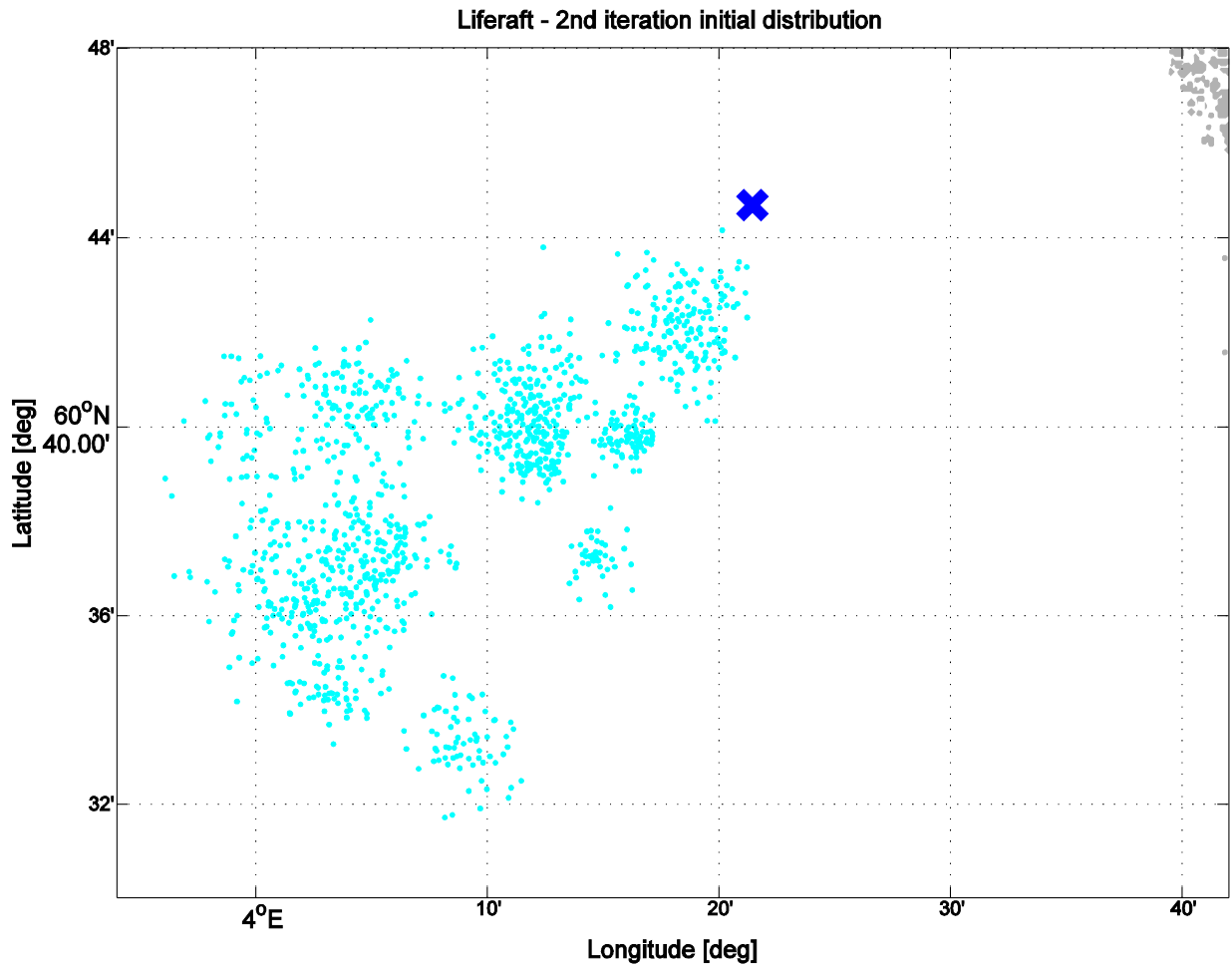

Figure 5, panel (a).

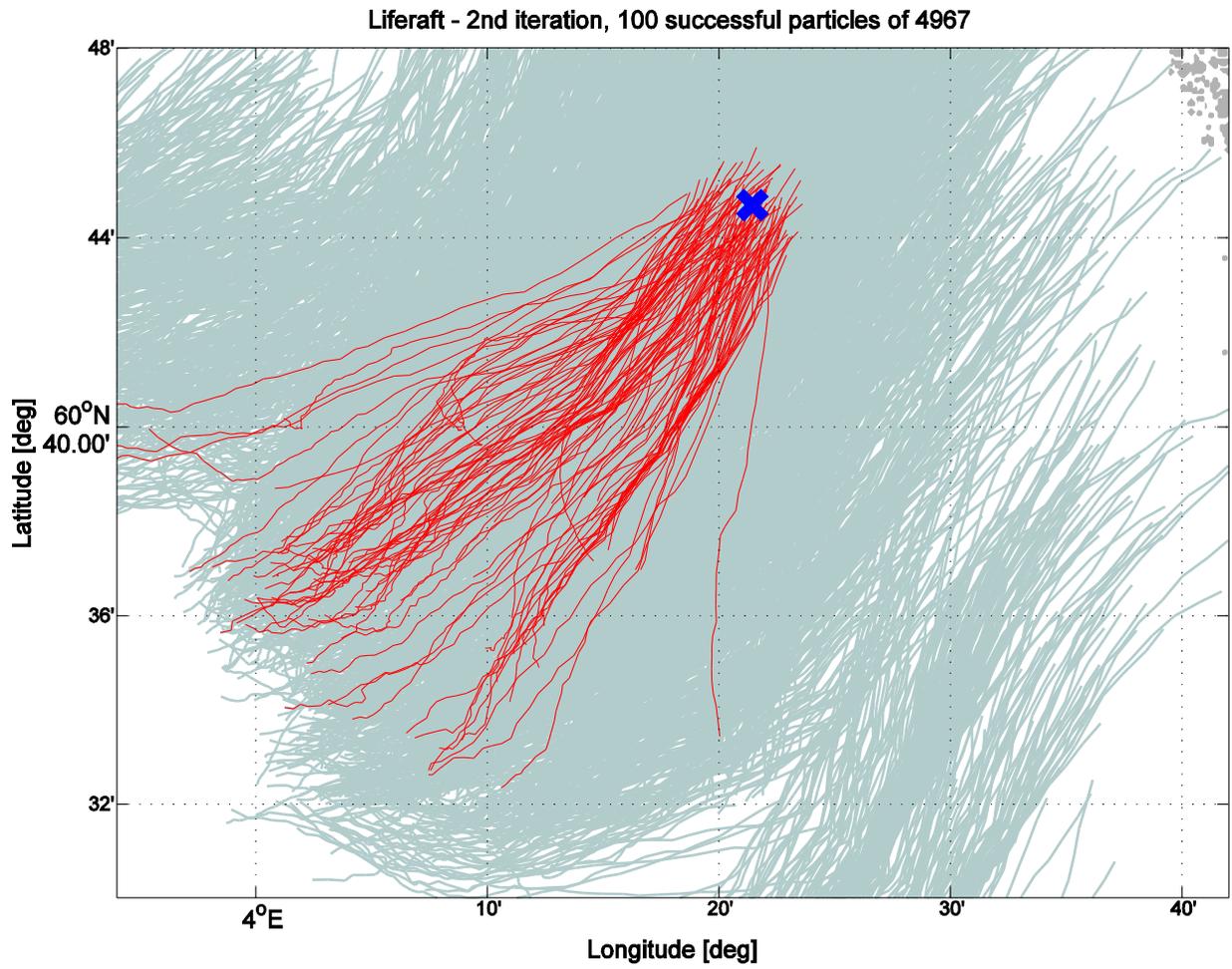

Figure 5, panel (b).

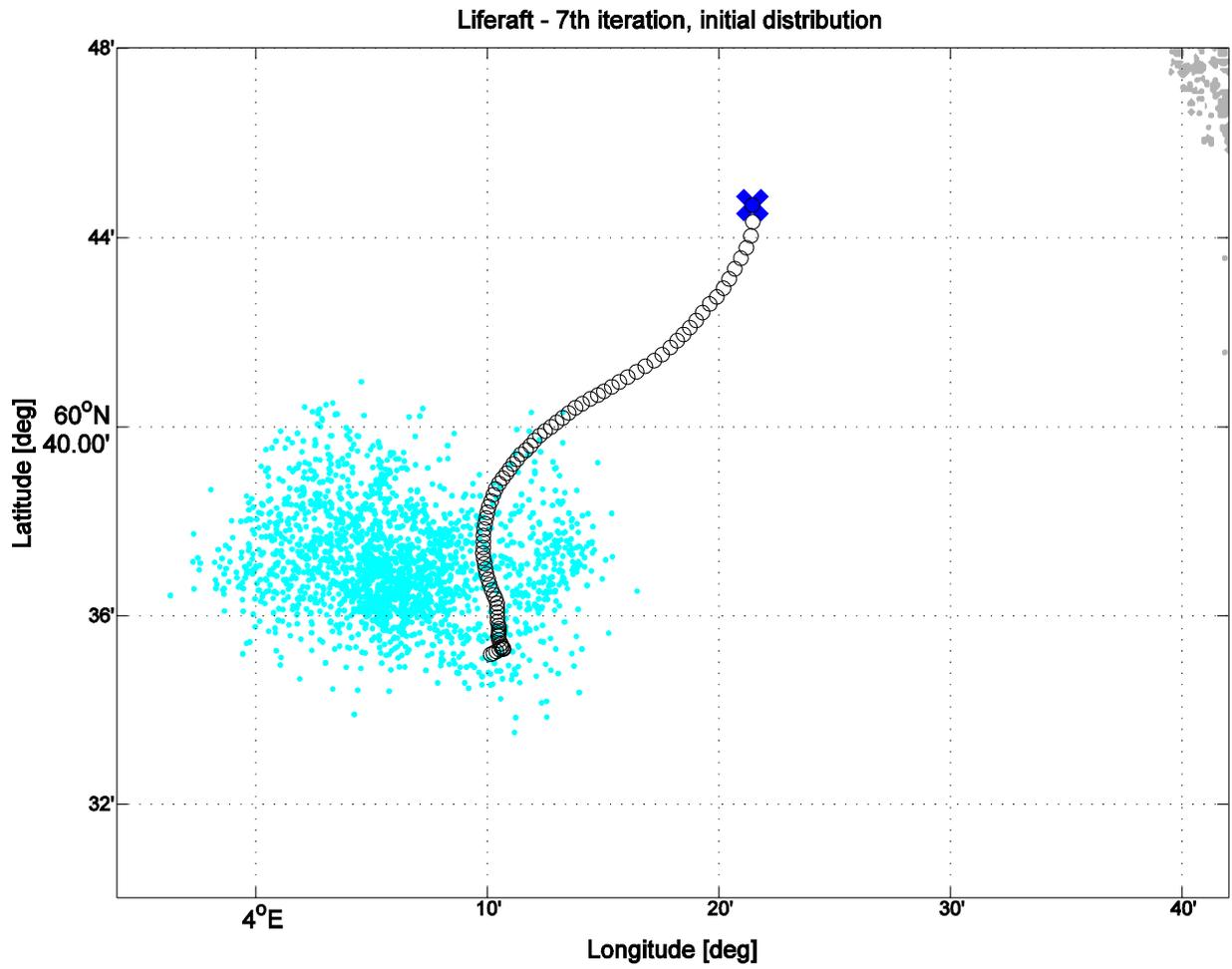

Figure 5, panel (c).

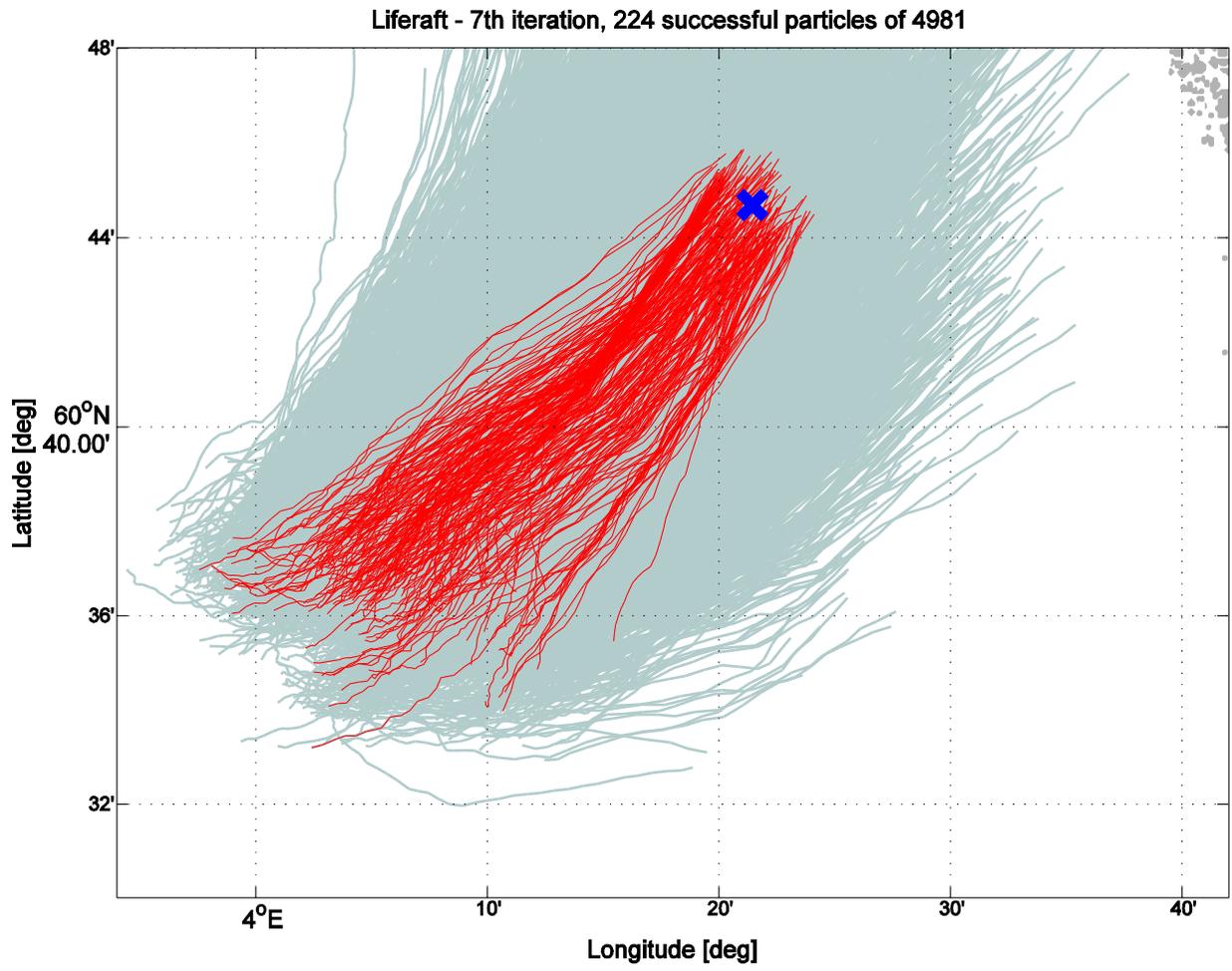

Figure 5, panel (d).

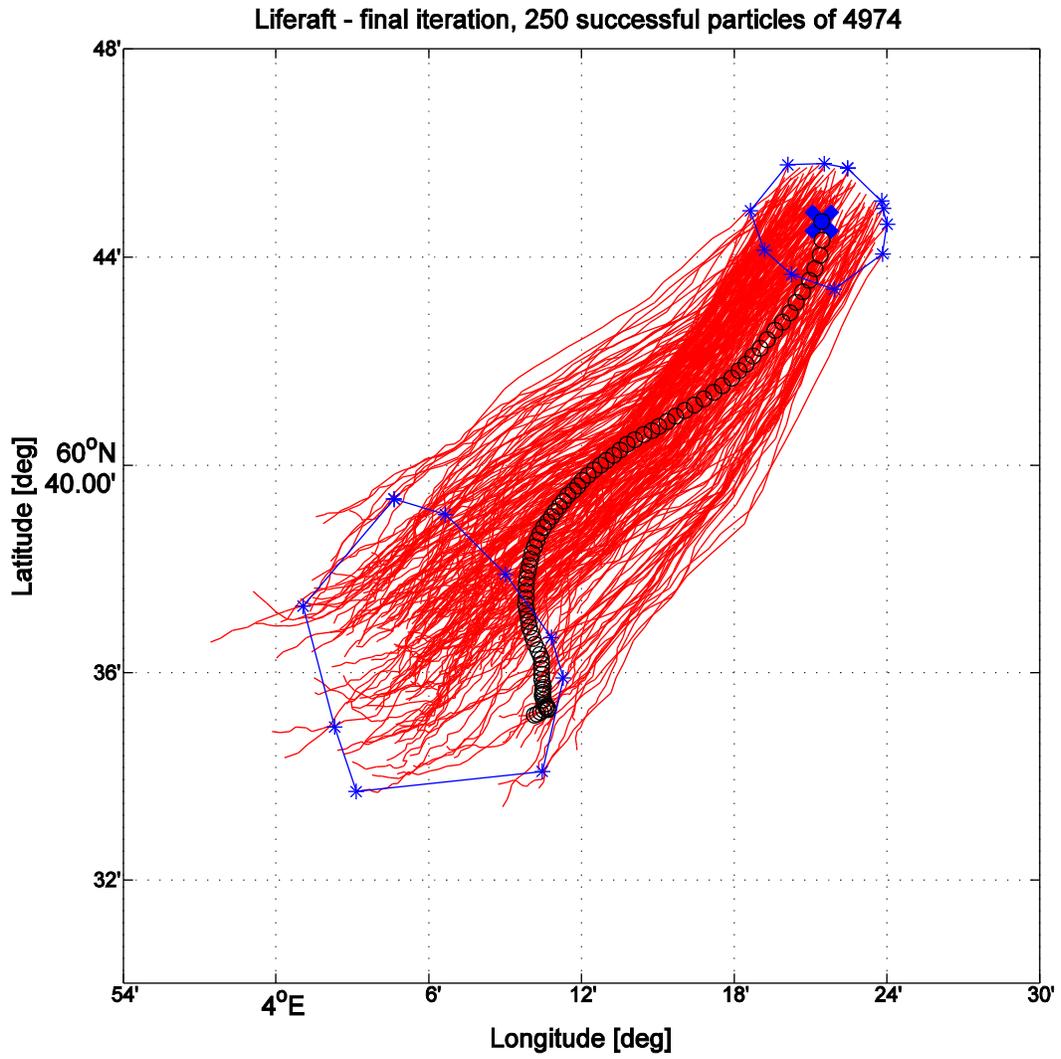

Figure 6.

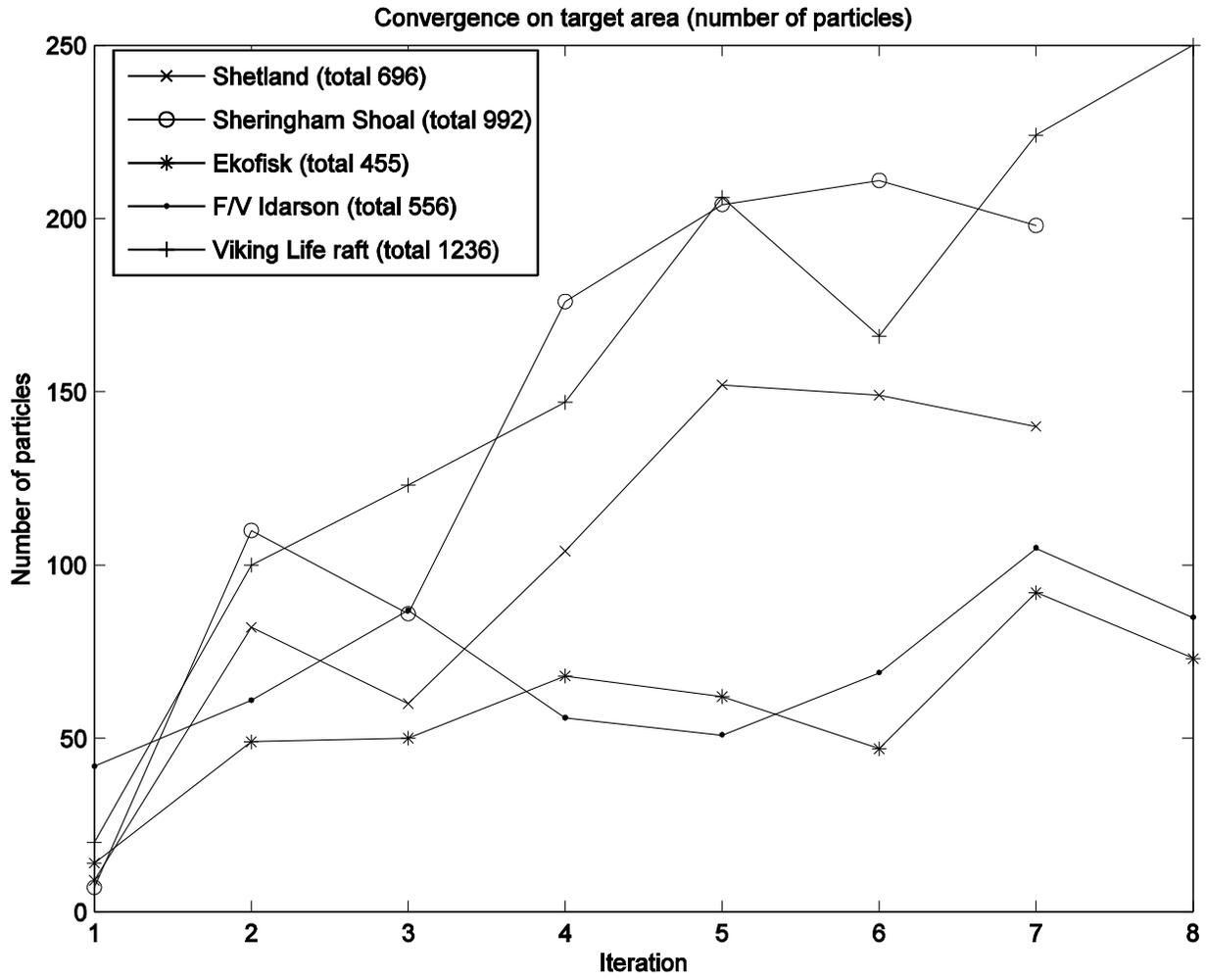

Figure 7.

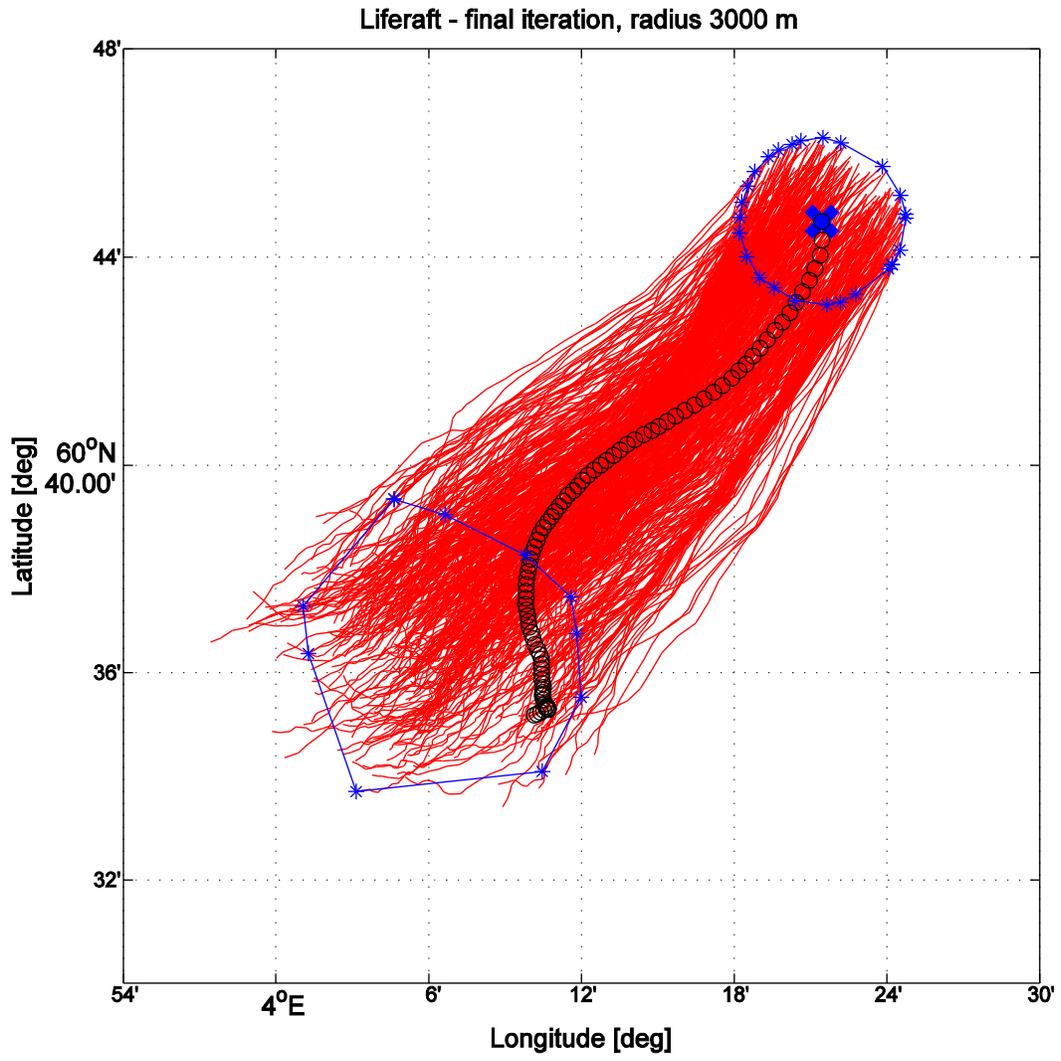

Figure 8.